\documentclass[10pt,conference]{IEEEtran}

\RequirePackage{ifthen}
\newboolean{debugLinks}\setboolean{debugLinks}{false}
\newboolean{debugCites}\setboolean{debugCites}{false}
\newboolean{debugPages}\setboolean{debugPages}{false}

\pdfpageattr{/Group << /S /Transparency /I true /CS /DeviceRGB >>}

\usepackage[utf8]{inputenc}
\usepackage[T1]{fontenc}
\usepackage{textcomp}
\usepackage{microtype}
\usepackage{calc}
\usepackage{tabularx}

\usepackage[normalem]{ulem}
\usepackage{footmisc}

\usepackage{lipsum}

\usepackage{silence}
\usepackage{csquotes}
\ifthenelse{\boolean{debugCites}}{
    \usepackage[backend=biber,style=verbose,bibstyle=ieee,doi=true,isbn=false,mincitenames=1,maxcitenames=2,minbibnames=1,maxbibnames=4]{biblatex}
}{
    \usepackage[backend=biber,style=ieee,bibstyle=ieee,doi=false,isbn=false,mincitenames=1,maxcitenames=2,minbibnames=1,maxbibnames=4]{biblatex}
}
\DeclareFieldFormat{sentencecase}{#1} 
\DeclareFieldFormat{titlecase}{#1} 
\usepackage{xpatch}
\xpatchbibmacro{textcite}{\addspace}{\addnbspace}{}{}
\xpatchbibmacro{Textcite}{\addspace}{\addnbspace}{}{}
\DefineBibliographyStrings{english}{
    andothers = et~al\adddot\addspace
}

\DeclareSourcemap{
    \maps[datatype=bibtex]{
        \map[overwrite=true]{
            \step[fieldsource=school,fieldtarget=institution]
        }
    }
}
\RenewDocumentCommand\cite{O{}m}{%
    \autocite[#1]{#2}
}

\usepackage{amsmath}
\usepackage{upgreek}

\usepackage{amssymb}
\usepackage{amsfonts}
\usepackage{amsthm}

\usepackage[range-phrase=--,per-mode=symbol-or-fraction,range-units=single,list-units=single,list-final-separator={\text{, and }},detect-all,forbid-literal-units=true]{siunitx}
\AtBeginDocument{
\DeclareSIUnit\bit{bit}
\DeclareSIUnit\byte{Byte}
\DeclareSIUnit\decibeli{dBi}
\DeclareSIUnit\decibelm{dBm}
\DeclareSIUnit\kmh{\kilo\meter\per\hour}
\DeclareSIUnit\mbps{\mega\bit\per\second}
\DeclareSIUnit\mph{mph}
\DeclareSIUnit\mw{\milli\watt}
\DeclareSIUnit\resourceblock{RB}
\DeclareSIUnit\vehicle{veh}
\DeclareSIUnit\watthour{Wh}
\DeclareSIUnit\usdollar{US\$}
\DeclareSIUnit\pplong{\acf{PP}}
\DeclareSIUnit\ppshort{\acs{PP}}
}

\usepackage{booktabs}
\usepackage[]{url}
\makeatletter
\def\url@cmsstyle{%
  \def\UrlSpecials{%
    \do\-{\mathchar`-}%
  }%
}
\makeatother

\usepackage[inline]{enumitem}

\usepackage[caption=false,font=footnotesize]{subfig}
\usepackage[pdftex]{graphicx}
\DeclareGraphicsExtensions{.pdf,.png,.jpg,.jpeg}
\usepackage{xfrac}

\usepackage{tikz}
\usetikzlibrary{arrows}
\usetikzlibrary{calc}
\usetikzlibrary{chains}
\usetikzlibrary{scopes}

\ifthenelse{\boolean{debugCites} \OR \boolean{debugLinks} \OR \boolean{debugPages}}{
    \WarningsOff[everypage]
    \usepackage[placement=top]{background}
    \SetBgContents{draft}
    \SetBgScale{1}
    \SetBgVshift{-7}
}{
}

\usepackage[american]{babel}
\usepackage{hyphenat}

\usepackage{algorithmic}
\usepackage{algorithm}

\ifthenelse{\boolean{debugLinks}}{
    \usepackage[breaklinks=true]{hyperref}
}{
}

\usepackage[capitalize,noabbrev,nameinlink]{cleveref}
\crefformat{footnote}{#2\textsuperscript{\normalfont #1}#3}
\usepackage{footmisc}

\RequirePackage{xstring}
\RequirePackage{xparse}
\RequirePackage[]{acro}

\makeatletter
\@ifpackagelater{acro}{2019/02/17}{
    \@ifpackagelater{acro}{2020/04/29}{
        \NewDocumentCommand\acrodef{mO{#1}mG{}}{\DeclareAcronym{#1}{short={#2}, long={#3}, #4}}
    }{
        \NewDocumentCommand\acrodef{mO{#1}mG{}}{\DeclareAcronym{#1}{short={#2}, long={#3}, foreign-plural={}, #4}}
    }
}{
    \NewDocumentCommand\acrodef{mO{#1}mG{}}{\DeclareAcronym{#1}{short={#2}, long={#3}, #4}}
}
\makeatother

\makeatletter
\@ifpackagelater{acro}{2020/04/29}{
    \NewAcroTemplate{inparen}{%
        \acroiffirstTF{%
            \acrowrite{long}%
            \acspace%
            \scalebox{0.5}[1]{(}%
            \acroifT{foreign}{\acrowrite{foreign}, }%
            \acrowrite{short}%
            \acroifT{alt}{ \acrotranslate{or} \acrowrite{alt}}%
            \acrogroupcite
            \scalebox{0.5}[1]{)}%
        }{%
            \acrowrite{short}%
        }%
    }
    \NewAcroTemplate{paren}{%
        \acroiffirstTF{%
            \acrowrite{long}%
            ,%
            \acspace%
            \acroifT{foreign}{\acrowrite{foreign}, }%
            \acrowrite{short}%
            \acroifT{alt}{ \acrotranslate{or} \acrowrite{alt}}%
            \acrogroupcite
        }{%
            \acrowrite{short}%
        }%
    }
    \NewAcroTemplate{inparenc}{%
        \acroiffirstTF{%
            \acrowrite{long}%
            ,%
            \acspace%
            \acroifT{foreign}{\acrowrite{foreign}, }%
            \acrowrite{short}%
            \acroifT{alt}{ \acrotranslate{or} \acrowrite{alt}}%
            \acrogroupcite
        }{%
            \acrowrite{short}%
        }%
    }
    \NewAcroTemplate{inparencc}{%
        \acroiffirstTF{%
            \acrowrite{long}%
            ,%
            \acspace%
            \acroifT{foreign}{\acrowrite{foreign}, }%
            \acrowrite{short}%
            \acroifT{alt}{ \acrotranslate{or} \acrowrite{alt}}%
            ,%
            \acrogroupcite
        }{%
            \acrowrite{short}%
        }%
    }
}{
    \DeclareAcroFirstStyle{inparen}{inline}
      {
        brackets = true,
        brackets-type = {%
          \scalebox{0.5}[1]{(}%
        }{%
          \scalebox{0.5}[1]{)}%
        }
      }
}
\makeatother

\usepackage{multirow}	
\usepackage{graphicx} 	
\acrodef{3GPP}{Third Generation Partnership Project}
\acrodef{5G}{5$^{\text{th}}$ Generation}
\acrodef{CNC}{Centralized Network Configuration}
\acrodef{C-V2X}{Cellular vehicle to everything}
\acrodef{DT}{Digital Twin}
\acrodef{ETSI}{European Telecommunications Standards Institute}
\acrodef{IETF}{Internet Engineering Task Force}
\acrodef{ITU-T}{International Telecommunication Union Telecommunication Standardization Sector}
\acrodef{KPI}{Key Performance Indicator}
\acrodef{MAC}{Medium Access}
\acrodef{MCS}{Modulation and Coding Scheme}
\acrodef{NDT}{Network Digital Twin}
\acrodef{OS}{Operating System}
\acrodef{PDR}{Packet Delivery Ratio}
\acrodef{PHY}{Physical}
\acrodef{RAN}{Radio Access Network}
\acrodef{RF}{Radio Frequency}
\acrodef{RSS}{Received Signal Strength}
\acrodef{SDN}{Software-Defined Networking}
\acrodef{SDO}{Standard Development Organization}
\acrodef{SINR}{Signal-to-Interference-plus-Noise Ratio}
\acrodef{SNR}{Signal-to-Noise Ratio}{short-indefinite={an}}
\acrodef{UDP}{User Datagram Protocol}

\newcommand{\useacronyms}{
}
\xapptocmd{\maketitle}{\useacronyms}{}{}
\xapptocmd{\acresetall}{\useacronyms}{}{}
\xapptocmd{\printbibliography}{\acuseall}{}{} 

\newcommand{\sectionacronyms}{
}
\AddToHook{cmd/section/before}{\sectionacronyms{}}

\usepackage{todonotes}

\usepackage[noadjust,noparboxrestore]{marginnote}
\makeatletter
\def\todo{%
    \begingroup%
    \color{magenta}%
    \ifnum\@floatpenalty<0\relax%
    \else%
        \setlength{\columnsep}{2cm} 
        \marginnote{\color{magenta}\rule{2pt}{1em}}%
        \obeylines%
        \obeyspaces%
        \begingroup\lccode`~=`\^^M\lowercase{\endgroup\def~}{\par\leavevmode}%
        \parindent0em%
        \catcode`\_=\active%
        \catcode`\<=\active\lccode`~=`<\lowercase{\def~}{$<$}%
        \catcode`\>=\active\lccode`~=`>\lowercase{\def~}{$>$}%
        \catcode`\#=\active\lccode`~=`\#\lowercase{\def~}{$\#$}%
        \catcode`\^=\active\lccode`~=`\^\lowercase{\def~}{$\hat{~}$}%
        \catcode`\&=\active\lccode`~=`\&\lowercase{\def~}{\&}%
    \fi%
    \todoCtd%
}\def\todoCtd#1{%
    TODO: #1%
    \ifx&#1&...\fi%
    \ifnum\@floatpenalty<0\relax%
    \else%
    \fi%
    \endgroup%
    \relax%
}
\makeatother

\NewDocumentCommand\IEEE{ s m >{\SplitArgument{4}{/}}d[] }{%
    \IfBooleanTF{#1}{}{IEEE\,}
    \nolinebreak[2]
    #2%
    \IfNoValueTF{#3}{%
    }{%
        \sommerIEEELettersSlashed#3%
    }%
}
\newcommand{\sommerIEEELettersSlashed}[5]{%
    \IfNoValueTF{#2}{%
    }{%
        \nolinebreak[3]
    }%
    #1%
    \IfNoValueTF{#2}{}{/#2}%
    \IfNoValueTF{#3}{}{/#3}%
    \IfNoValueTF{#4}{}{/#4}%
    \IfNoValueTF{#5}{}{/#5}%
}

\NewDocumentCommand\sommerInvisibleText{ m }{%
    {
    \fontsize{.1}{.1}\selectfont
    \pdfliteral page{q 3 Tr}
    #1%
    \pdfliteral page{Q}
    }
}

\NewDocumentCommand\sommerMultiUrl{ d[] >{\SplitArgument{4}{,}}m d[] }{%
    \texttt{%
    \def\sommerMultiUrlFront{\IfNoValueTF{#1}{}{#1}}%
    \def\sommerMultiUrlEnd{\IfNoValueTF{#3}{}{#3}}%
    \def\sommerMultiUrlSpace{\;}%
    \sommerMultiUrlMiddle#2%
    }%
}
\newcommand{\sommerMultiUrlMiddle}[5]{%
    \sommerMultiUrlFront{}%
    \sommerMultiUrlSpace{}\{\sommerMultiUrlSpace{}%
    \IfNoValueTF{#1}{}{\sommerInvisibleText{\sommerMultiUrlFront{}}#1}%
    \IfNoValueTF{#2}{}{\sommerInvisibleText{\sommerMultiUrlEnd{}}\sommerMultiUrlSpace{},\sommerMultiUrlSpace{}\sommerInvisibleText{\sommerMultiUrlFront{}}#2}%
    \IfNoValueTF{#3}{}{\sommerInvisibleText{\sommerMultiUrlEnd{}}\sommerMultiUrlSpace{},\sommerMultiUrlSpace{}\sommerInvisibleText{\sommerMultiUrlFront{}}#3}%
    \IfNoValueTF{#4}{}{\sommerInvisibleText{\sommerMultiUrlEnd{}}\sommerMultiUrlSpace{},\sommerMultiUrlSpace{}\sommerInvisibleText{\sommerMultiUrlFront{}}#4}%
    \IfNoValueTF{#5}{}{\sommerInvisibleText{\sommerMultiUrlEnd{}}\sommerMultiUrlSpace{},\sommerMultiUrlSpace{}\sommerInvisibleText{\sommerMultiUrlFront{}}#5}%
    \sommerInvisibleText{\sommerMultiUrlEnd{}}%
    \sommerMultiUrlSpace{}\}\sommerMultiUrlSpace{}%
    \sommerMultiUrlEnd{}%
}

\usepackage{listings}
\usepackage{tcolorbox}
\usepackage[table, svgnames,dvipsnames]{xcolor}

\usepackage{pifont}

\makeatletter
\newcommand\mynobreakpar{\par\nobreak\@afterheading}
\makeatother

\csname endofdump\endcsname
\usepackage{wrapfig}

\usepackage{tabularray}

\usepackage[table]{xcolor}

\usepackage{makecell}
\usepackage{csquotes}
\usepackage{fancyhdr}
\fancypagestyle{firstpage}{%
  \fancyhf{} 

  \fancyfoot[L]{\scriptsize
    This paper has been published in the Proceedings of the IEEE Future Networks World Forum (FNWF), 2025.\\
    \copyright{} 2025 IEEE. DOI: 10.1109/FNWF66845.2025.11317602.
  }
}

\title{User-Centric Comparison of 5G NTN and DVB-S2/RCS2 Using OpenAirInterface~and~OpenSAND}

\author{
    \IEEEauthorblockN{Sumit Kumar, Juan Carlos Estrada-Jimenez, Ion Turcanu} \\
    Luxembourg Institute of Science and Technology, Luxembourg \\
    \{sumit.kumar, juan.estrada-jimenez, ion.turcanu\}@list.lu
}

\begin{document}

\maketitle
\thispagestyle{firstpage} 

\begin{abstract}
The integration of satellite networks into next-generation mobile communication systems has gained considerable momentum with the advent of 5G Non-Terrestrial Networks (5G-NTN). Since established technologies like DVB-S2/RCS2 are already widely used for satellite broadband, a detailed comparison with emerging 5G NTN solutions is necessary to understand their relative merits and guide deployment decisions. This paper presents a user-centric, end-to-end evaluation of these technologies under realistic traffic conditions, showing how differences in architecture and protocols impact application-layer performance.
Utilizing the 6G Sandbox platform, we employ OpenAirInterface to emulate 5G NTN and OpenSAND for DVB-S2/RCS2, replicating transparent payload GEO satellite scenarios under uniform downlink conditions. A range of real-world applications, such as web browsing, file downloads, and video streaming, are tested across both systems and systematically analyzed. While the emulation lacks real-time capability, it reveals key strengths and limitations of each approach, helping identify suitable deployment scenarios for 5G NTN and DVB-S2/RCS2.
\end{abstract}

\section{Introduction}\label{sec:intro}
Satellite communication systems are increasingly important for extending broadband to underserved regions. 3GPP added Non-Terrestrial Network (NTN) support in Release-17 with enhancements in Release-18 and ongoing studies toward Release-19, while DVB-S2 and its return-channel extension DVB-RCS2 remain widely used for satellite broadband. Prior comparisons of 5G-NTN and DVB-S2/RCS2 typically focus on PHY/MAC metrics and often rely on proprietary tool-chains, limiting reproducibility and leaving application-layer, user-centric performance underexplored.

This paper presents an end-to-end, application-layer comparison of 5G-NTN and DVB-S2/RCS2 in a transparent-payload GEO scenario using open-source platforms: OpenAirInterface5G-NTN \cite{kumar2022openairinterface} and OpenSAND \cite{opensand}, deployed on the 6G Sandbox \cite{6gsandbox}. We evaluate user-facing KPIs (jitter, file download time, video start time, webpage load time) to assess practical implications of protocol and architectural differences. While hardware-based emulation provides higher fidelity, simulator-based testbeds offer reproducible, cost-effective environments for early-stage NTN evaluation.

Our main contributions are: (i) an open, reproducible testbed comparison using OAI and OpenSAND; (ii) an end-to-end, application-layer KPI analysis; and (iii) identification of software/platform limitations relevant for realistic NTN experiments.

\section{Prior Works}\label{sec:sota}
Comparative studies of 5G NR NTN and DVB-S2/RCS2 have produced useful PHY/MAC insights but tend to share two limitations: reliance on non-public toolchains and little attention to application-level user experience. For instance, Sormunen et al. \cite{sormunen2025simulative} and Huikko \cite{huikko2023comparison} present ns-3–based system/link comparisons that emphasize spectral efficiency and throughput but use simulators or setups that are not publicly reproducible. George et al. \cite{george20215g} investigate RF impairments and show waveform sensitivity (e.g., higher degradation for 5G NR under amplifier nonlinearity), while Delbeke \cite{delbeke2022satellite} focuses on return-link waveform efficiency and ModCod trade-offs. The ETSI technical comparison \cite{etsiTR103886} offers a valuable standards-level analytical baseline but lacks implementation-level validation and empirical, end-to-end measurements. 

Complementary to simulator-based comparisons, 5G-NTN OAI based prototypes have validated GEO over-the-air and in-lab feasibility and explored LEO adaptations \cite{kumar20225g, volk20245g, kumar20235g, kumar20225ag}. These efforts focused on protocol adaptations and demonstrator readiness rather than application-layer KPI benchmarking across technologies, which is the focus of our work.

Collectively, these works advance link-level understanding (SE, SINR, PAPR, BLER) but rarely evaluate how protocol and architectural differences translate to application-layer KPIs (jitter, download time, video startup, webpage load) under realistic traffic. They also provide limited discussion on simulation realism, timing/pacing effects in software emulators, and reproducible configuration details — all of which are important when comparing systems with fundamentally different framing and return-link behavior. This motivates the need for open, reproducible, end-to-end experiments focused on user-centric metrics, which is the objective of the present study.

\section{DVB-S2/RCS2 and 5G-NTN: Technologies and Tools}

\noindent \textbf{DVB-S2/RCS2.} DVB-S2 (and S2X) together with DVB-RCS2 form an established satellite broadband stack originally developed for broadcasting and extended for interactive services. Forward links use DVB-S2/S2X, while DVB-RCS2 provides a scheduled return channel; IP traffic is carried with Generic Stream Encapsulation (GSE), which provides fragmentation, reassembly, and multiplexing and is well suited to broadcast/TDM delivery.

\noindent \textbf{5G-NTN.} In contrast, 5G-NTN (3GPP Rel-17+) provides a native, end-to-end IP-centric protocol stack (RLC/PDCP/SDAP, GTP-U, etc.) designed for bidirectional mobile services. It natively handles fragmentation, QoS, and bearer-based tunnelling, and its flexible scheduling and bearer model make it better suited to interactive, low-latency IP traffic over satellite compared to GSE/TDM-based approaches.

\subsection*{Emulation Platforms}
\noindent \textbf{OpenSAND.} OpenSAND, developed by CNES and Thales Alenia Space \cite{opensand}, is a satellite-network emulator that reproduces key PHY and system-level characteristics such as propagation delay, link attenuation, and payload models (gateway, transparent payload, terminal). It supports configurable forward/return links, delay profiles, and attenuation/clear-sky parameters, enabling end-to-end validation of satellite protocols and services without requiring RF hardware.

\noindent \textbf{OpenAirInterface5G-NTN.} OpenAirInterface5G-NTN is an open-source, 3GPP-compliant framework that extends OAI’s terrestrial 5G stack to NTN conditions \cite{kumar2022openairinterface}. It incorporates satellite-specific enhancements (large Doppler, long one-way delays, transparent or regenerative payloads), implements gNB and UE functions across PHY/MAC/RLC/PDCP, and exposes modular configuration for orbital parameters, link models, and scheduling—making it well suited for research, prototyping, and experiment-driven evaluation of satellite-integrated 5G systems.

\section{Real-Time Issues}
\noindent In simulation environments such as OpenAirInterface’s RF simulator and OpenSAND, the system operates in a non-real-time mode, which introduces several critical challenges. When hardware is present \cite{kumar2023experimental}, the processing of samples is inherently paced by the physical sampling rate—for instance, a typical 5G NR system operating at a sampling rate of approximately 7.68 MSPS processes exactly that many samples per second. This hardware-driven pacing ensures predictable timing behavior and consistent real-time execution. However, in the absence of hardware, the host operating system processes the samples without strict timing constraints. This leads to non-deterministic scheduling and variable processing rates. The lack of precise timing guarantees also affects interactions between protocol layers; for example, real-time signaling queues and interrupt-driven processing at the PHY and MAC layers are replaced by software polling or artificial delays. As a result, latency measurements such as round-trip time (RTT), jitter, and throughput become imprecise and inconsistent. These factors cause the protocol stack behavior in simulation to deviate from that of a real RF environment, complicating the interpretation of timing-sensitive results.

\noindent Despite these limitations, it remains feasible and meaningful to perform comparative measurements between 5G (using OpenAirInterface) and DVB-S2/RCS2 (using OpenSAND) under the same simulation conditions. Both simulations run on identical CPU resources, ensuring that processing capacity and system load are equivalent. Maintaining consistent computational environments allows for fair relative performance comparisons between the two systems, isolating differences attributable to protocol and system design rather than hardware variability.

\section{Experiment Setup}
\subsection{Hardware Set-up}
\begin{figure}[ht]
    \centering
    \includegraphics[width=\columnwidth]{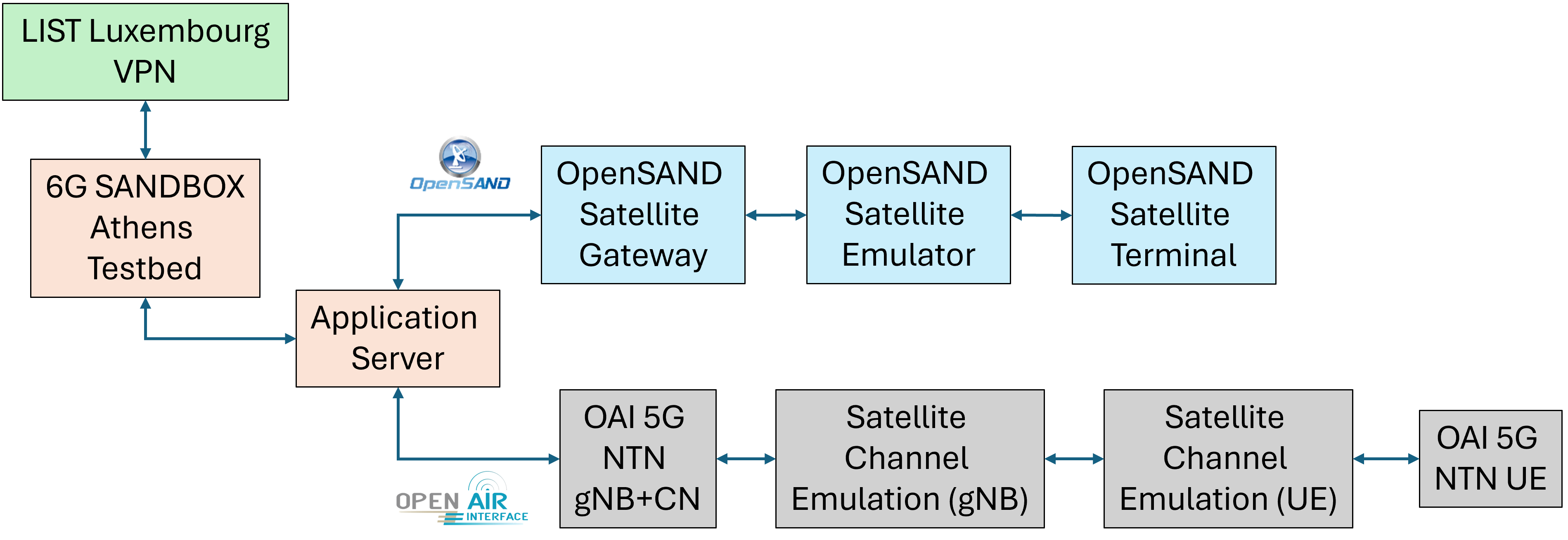}
    \caption{Experiment set-up at the 6G-Sandbox Platform.}
    \label{fig:exp-set-up}
\end{figure}

\noindent All experiments were conducted on the Athens 6G Sandbox platform \cite{6gsandbox}, which provides a controlled environment for end-to-end experimentation with satellite network components. Our set-up, as illustrated in \Cref{fig:exp-set-up}, each node (OAI gNB, OAI UE, openSAND satellite emulator, openSAND gateway, openSAND terminal) is equipped with an \textit{Intel Xeon E5-2680 v3} processor and dedicated 8 physical cores operating at 2.50~GHz.
The set-up spans two main environments: OAI-based 5G NTN and the openSAND-based DVB-S2/RCS2. The openSAND chain includes a satellite gateway, transparent payload GEO channel emulator, and terminal. The OAI setup includes a 5G NTN gNB, Core network, and a UE, configured with satellite delay parameters to mirror the openSAND link. A common application server is connected to both environments and used to generate traffic and collect logs. All traffic flows through the respective satellite-emulated paths, enabling side-by-side performance evaluations under controlled and reproducible conditions.

\subsection{Software Set-up}
\noindent To ensure a fair and equivalent comparison between the 5G NTN waveform (via OpenAirInterface) and the DVB-S2/RCS2 waveform (via OpenSAND), both systems are configured under identical network conditions, traffic profiles, and satellite parameters.  The following subsections outline the key configuration choices made to align both systems as closely as possible within the constraints of their respective implementations.

\subsubsection{Bandwidth Configuration}
Both OAI and OpenSAND are configured to use a downlink bandwidth of 5~MHz. In OAI, the bandwidth is set by adjusting the number of Physical Resource Blocks (PRBs). In our configuration, the subcarrier spacing (SCS) is set to 15~kHz, so by choosing the number of PRBs as 25, the total bandwidth can be approximated as:
\[
\text{Bandwidth} = 12 \times \text{PRB} \times \text{SCS} = 12 \times 25 \times 15\,\text{kHz} = 4.5\,\text{MHz}
\]
Considering guard bands and implementation overhead, this configuration roughly corresponds to a nominal 5~MHz channel bandwidth.
\\
\noindent In \texttt{OpenSAND}, which is based on a single-carrier DVB-S2 waveform, the system bandwidth is configured using the symbol-rate and the roll-off factor. These two parameters jointly determine the \textit{occupied bandwidth} of the transmitted signal. The roll-off factor ($\alpha$) defines the excess bandwidth beyond the Nyquist rate to accommodate practical pulse shaping and filtering. The total occupied bandwidth \(B\) is given by \(B = R_s(1+\alpha)\), where \(R_s\) is the symbol rate and \(\alpha\) is the roll-off factor. To approximately match an effective 5~MHz bandwidth in our setup, we selected the $R_s = 5 \text{Msps}$ and $\alpha = 0.35$, as shown in the configuration below: 
{\small
\begin{verbatim}
<roll_off>
<forward>0.350000</forward>
</roll_off>
<symbol_rate>
5000000.000000
</symbol_rate>
\end{verbatim}
}

\noindent This configuration sets a forward link roll-off of 0.35, resulting in an occupied bandwidth \(B = 5\ \mathrm{Msps}\times(1+0.35)=6.75\ \mathrm{MHz}\). Although this exceeds the nominal 5~MHz, it accounts for the spectral shaping required by the DVB-S2 waveform. 

\subsubsection{Satellite Payload and Delay}
A GEO satellite configuration is used for both systems, introducing a typical Round-Trip-Time(RTT) of approximately 520~ms (one-way propagation delay $\approx$ 260~ms). Both platforms are configured with a \textit{transparent payload} model (satellite acts as a bent-pipe relay without any onboard signal processing). 

In OAI, the transparent satellite payload and required propagation delay are configured in \texttt{servingCellConfigCommon} inside the configuration file as follows: 
\begin{itemize}[leftmargin=*]
    \item \texttt{cellSpecificKoffset\_r17}: Scheduling offset in slots for 15~kHz SCS. A typical value is 520.
    \item \texttt{ta-Common-r17}: Timing advance to compensate for the one-way propagation delay. The granularity is $4.072 \times 10^{-3}~\mu$s:
    \[
    \texttt{ta-Common-r17} = \frac{260 \times 10^3~\mu\text{s}}{4.072 \times 10^{-3}~\mu\text{s}} \approx 63813480
    \]
    \item Ephemeris data: Satellite position and velocity vectors in the ECEF (Earth-Centered, Earth-Fixed) frame:
    \begin{itemize}[leftmargin=*]
        \item \texttt{positionX-r17 = 0}, \texttt{positionY-r17 = 0}, \texttt{positionZ-r17 = 27527692} \\
        (This corresponds to a 35,786~km GEO altitude with a step size of 1.3~m.)
        \item \texttt{velocityVX/VY/VZ-r17 = 0} \\
        (For GEO satellites, velocity is negligible in the ECEF frame.)
    \end{itemize}
\end{itemize}
\noindent The final configuration snippet we used is:
{\small
\begin{verbatim}
cellSpecificKoffset_r17 = 520;
ta-Common-r17           = 63813480;
positionX-r17           = 0;
positionY-r17           = 0;
positionZ-r17           = 27527692;
velocityVX-r17          = 0;
velocityVY-r17          = 0;
velocityVZ-r17          = 0;
\end{verbatim}
}
Besides, for the one-way delay of 260~ms, the following arguments are added to the command lines for both gNB and UE
\begin{verbatim}
--rfsimulator.prop_delay 260
\end{verbatim}

Similarly, OpenSAND allows selection of the transparent payload type through the profile and topology configuration of the emulator, gateway, and terminal. In our setup, a GEO satellite with a transparent payload and a one-way delay of 260~ms is emulated, matching the OAI configuration. In this case, a transparent payload is configured for both the forward and return links as follows:
{\small
\begin{verbatim}
<assignments>
<forward_regen_level>
Transparent
</forward_regen_level>
<return_regen_level>
Transparent
</return_regen_level>
</assignments>
\end{verbatim}
}
Furthermore, to emulate GEO satellite delay, a constant one-way delay of 260~ms is set in the profile configuration:
{\small
\begin{verbatim}
<delay>
<delay_type>ConstantDelay</delay_type>
<delay_value>260</delay_value>
</delay>
\end{verbatim}
}

\subsection{Modulation and Coding Scheme Configuration}
In our work, a fixed Modulation and Coding Scheme (called MCS for OAI and ModCod for openSAND) is used across both platforms to ensure deterministic and repeatable testing conditions. Manually selecting static MCS/ModCod enables controlled evaluation of protocol and delay impacts without introducing variability from adaptive link adaptation mechanisms. Besides, we configured both systems (OAI and openSAND) to operate under the lowest SNR levels they support (see \cref{sec:snr}). Accordingly, the lowest available MCS/ModCod levels are selected to ensure robust operation. Although the spectral efficiency (SE) of the minimum MCS in 5G NR differs from that of the lowest ModCod in DVB-S2/RCS2, the selected values in each system represent the most conservative and noise-resilient setting allowed by their respective standards. This provides a common baseline for performance comparison.

In OAI, the MCS is controlled through the maximum MCS value defined in the gNB configuration file. The following configuration sets the uplink and downlink MCS to 1 (QPSK, Coding Rate 0.0762 and SE 0.1524 bps/Hz): 
\begin{verbatim}
dl_max_mcs = 1; ul_max_mcs = 1;
\end{verbatim}

In openSAND, the ModCod is fixed by selecting the waveform index in the gateway and terminal profiles using the \texttt{<wave\_form>} parameter. The following setting sets the forward and return link ModCod to 1 (QPSK, Coding Rate 0.2 and SE of 0.4 bps/Hz)
\begin{verbatim}
<wave_form>1-1</wave_form>
\end{verbatim}

\noindent \textit{Please note that although the spectral efficiencies and coding rates differ, both configurations represent the most robust, noise-tolerant settings in their respective standards, offering a stable baseline for comparison under low-SNR conditions.}

\subsection{SNR Configuration}\label{sec:snr}
\noindent To ensure a fair and robust comparison between 5G-NTN and DVB-S2/RCS2 under marginal link conditions, simulations were conducted using a “barely surviving” link setup, where decoding just succeeds with minimal margin. Given that tools like OAI’s RF Simulator and openSAND lack fully calibrated power and noise models, absolute signal and noise levels are abstracted. Instead, SNR was tuned on each platform to enable end-to-end link establishment without enforcing strict performance thresholds. This approach offers a practical and reproducible baseline that highlights protocol robustness under typical low-SNR satellite conditions.

In the OAI RF simulator, we empirically tuned the channel noise power. Starting from 0dB (low noise), we gradually increased it until connectivity failed. A value of +3dB marked the threshold—beyond it, decoding consistently failed, while at +3dB, the system remained just operational. Because the simulator’s signal power is normalized to 0 dB, a noise power of +3 dB yields an effective SNR of –3 dB, where signal and noise are comparable. The parameter is set through rfsimu\_setchanmod\_cmd in the RF-simulator source file \cite{openairinterface5g_rfsimulator}.

In OpenSAND, the parameter \texttt{ideal\_attenuation\_value} controls the SNR. It is applied identically at both the gateway and the terminal. From experimentation and trials, we determined that an attenuation value of 44 dB provides the desired threshold. The XML snippet below shows the uplink and downlink configuration in our setup:
{\small
\begin{verbatim}
<uplink_attenuation>
<clear_sky>50.000000</clear_sky>
<attenuation_type>Ideal</attenuation_type>
<ideal_attenuation_value>44
</ideal_attenuation_value>
</uplink_attenuation>
<downlink_attenuation>
<clear_sky>50.000000
</clear_sky>
<attenuation_type>Ideal</attenuation_type>
<ideal_attenuation_value>44
</ideal_attenuation_value>
</downlink_attenuation>
\end{verbatim}
}
The \texttt{clear\_sky} parameter defines the baseline SNR(dB) when no attenuation is applied to the link. In our configuration, this value is set to \(50\,\text{dB}\), representing the ideal link condition without any channel loss. The \texttt{ideal\_attenuation\_value} specifies the attenuation applied to the signal in dB. Through iterative tuning, we found that an attenuation of \(44\,\text{dB}\) for both uplink and downlink yields the minimum SNR at which reliable decoding is still possible. This value was determined by gradually increasing attenuation until the link reached its viability limit. At this point, the instantaneous SNR for each link leg is computed as~\cite{opensand_link_budget}:
\[
\text{SNR}_{\text{link}} = \text{clear\_sky} - \text{attenuation} = 50\,\text{dB} - 44\,\text{dB} = 6\,\text{dB}
\]

\begin{table}[ht]
\centering
\caption{Experiment Parameters for OAI and openSAND}
\label{tab:experiment_params}
\begin{tabularx}{\columnwidth}{|l|X|X|}
\hline
\textbf{Parameters} & \textbf{OAI 5G NTN} & \textbf{DVB-S2/RCS2} \\
\hline
Satellite Type & GEO & GEO \\
\hline
Satellite Payload & Transparent & Transparent \\
\hline
RTT & $\approx$ 520 ms & $\approx$ 520 ms \\
\hline
Link SNR & -3 dB & 6 dB \\
\hline
MCS (AMC Disabled) & MCS-1 & MODCOD-1 \\
\hline
Downlink Bandwidth & 5MHz & 5MHz \\
\hline
Protocol-based retransmissions & Disabled & Disabled \\
\hline
Application layer retransmissions & Enabled & Enabled \\
\hline
Scheduler & Default & Default \\
\hline
Traffic & IP based & IP based \\
\hline
CPU Cores & 8 (gNB and UE each) & 8 (Emulator, Gateway, and Terminal each) \\
\hline
\end{tabularx}
\end{table}
\Cref{tab:experiment_params} summarizes all the settings used for our experiments. 

\subsection{Traffic Generation and Measurement}
To evaluate and compare the performance of 5G NTN and DVB-S2/RCS2 links under realistic application scenarios, we employ the following lightweight open-source tools focused on application-layer traffic generation and logging.
\begin{itemize}[leftmargin=*, labelsep=0.5em]
    \item \texttt{ping}: To measure RTT and Jitter. 
    \item \texttt{curl}: To emulate application-layer interactions such as file downloads, video streaming, and web page access. 
    \item \texttt{wget}: For file downloads.
\end{itemize}
Using these tools, downlink traffic is generated at the application server (see \Cref{fig:exp-set-up}) and routed through the system-specific interfaces of OAI (\texttt{oaitun\_ue1}) and openSAND (\texttt{opensand\_br}). Traffic flows follow a client-server model, where the client resides behind the simulated UE/user terminal and the server is hosted on the remote application server. This setup mimics realistic Internet-based communication patterns and enables end-to-end performance evaluation.

\section{Experiments \& Results}
\noindent This section presents a series of experiments designed to compare the performance of 5G NR NTN (via OAI) and DVB-S2/RCS2 (via openSAND) under identical conditions. The goal is to assess how differences in the protocol stack and system architecture impact user-centric application performance across a range of scenarios.

\noindent \textbf{Disclaimer:} Experiments were conducted in non-real-time simulation without hardware pacing, so absolute metrics (jitter, delay, throughput) may not match real deployments. Still, both systems ran under identical conditions (hardware, delay, and traffic), making the \emph{relative performance differences} valid for comparing protocol behavior.

\subsection{Experiment-1: Jitter Characterization}
\noindent This test aims to characterize packet jitter by measuring variations in delay over a sequence of network transmissions. A simple ping-based method was used, sending 100 ICMP echo requests from the application server to (a) the 5G-NR NTN UE and (b) the DVB Terminal. RTTs were recorded, and jitter was computed as the average absolute difference between consecutive RTTs. This metric reflects delay variability, which is crucial for real-time and streaming application performance.

\begin{table}[ht]
\centering
\caption{Jitter Comparison (ms)}
\label{tab:jitter-comparison}
\begin{tabular}{|c|c|c|c|}
\hline
\textbf{\#} & \textbf{5G-NTN} & \textbf{DVB-S2/RCS2} & \textbf{DVB / 5G Ratio} \\
\hline
1 & 4.15 & 13.60 & 3.28 \\
\hline
2 & 3.76 & 13.57 & 3.61 \\
\hline
3 & 3.67 & 11.83 & 3.22 \\
\hline
4 & 4.29 & 12.06 & 2.81 \\
\hline
5 & 4.13 & 12.63 & 3.06 \\
\hline
\end{tabular}
\end{table}

\subsection*{Result \& Discussion}
As shown in \Cref{tab:jitter-comparison}, the 5G-NTN link demonstrates significantly lower and more stable jitter (mean $\approx$ 4ms) compared to DVB-S2/RCS2 (mean $\approx$ 12.7ms). This disparity stems from architectural and protocol stack differences. The OAI 5G-NTN implementation uses a fully integrated 3GPP stack comprising RLC, PDCP, and SDAP layers, which enable efficient buffering, reordering, and flow control—resulting in smoother packet inter-arrival times and reduced delay variability. Additionally, its built-in scheduling and bearer-based traffic handling enhance temporal consistency.

In contrast, DVB-S2/RCS2, as emulated by openSAND, employs the GSE protocol over the DVB-S2 PHY layer. GSE multiplexes IP datagrams into baseband frames in batch mode, introducing serialization and fragmentation delays. This batch framing, combined with low ModCod settings, increases inter-packet timing variance due to longer transmission intervals. Furthermore, DVB-S2/RCS2 typically uses time-division multiplexing (TDM), which under low-throughput conditions leads to bursty packet arrivals and queuing delays. By contrast, OAI operates in an FDD mode that provides continuous channel availability, contributing to more stable packet delivery and reduced jitter.

\subsection{Experiment-2: Video Load Time}
\noindent This experiment measures the Time To First Frame (TTFF) to assess video startup delay. The UE requests a video file from the application server running a Python-based HTTP server. A \texttt{curl} based tool captures key timing metrics—\texttt{TTFF}, \texttt{time\_connect}, and \texttt{time\_starttransfer} to determine the time until the first video byte is received. This approach effectively evaluates video startup latency, a key factor in streaming quality.

\begin{table}[ht]
\centering
\caption{Video Load Time}
\label{tab:video-nofile}
\begin{tabular}{|>{\centering\arraybackslash}p{0.13cm}
                |>{\centering\arraybackslash}p{0.65cm}
                |>{\centering\arraybackslash}p{0.95cm}
                |>{\centering\arraybackslash}p{0.85cm}
                |>{\centering\arraybackslash}p{0.75cm}
                |>{\centering\arraybackslash}p{0.95cm}
                |>{\centering\arraybackslash}p{0.85cm}|}
\hline
\textbf{\#} & \multicolumn{3}{c|}{\textbf{5G-NTN (s)}} & \multicolumn{3}{c|}{\textbf{DVB-S2/RCS2 (s)}} \\
\hline
 & TTFF & Connect & Start Transfer & TTFF & Connect & Start Transfer \\
\hline
1 & 55.09 & 0.52 & 1.06 & 114.17 & 0.53 & 1.05 \\
\hline
2 & 60.09 & 0.52 & 1.06 & 114.11 & 0.51 & 1.04 \\
\hline
3 & 55.65 & 0.52 & 1.06 & 114.21 & 0.53 & 1.06 \\
\hline
4 & 54.96 & 0.52 & 1.07 & 115.49 & 0.53 & 1.06 \\
\hline
5 & 54.85 & 0.52 & 1.06 & 114.11 & 0.54 & 1.06 \\
\hline
\end{tabular}
\end{table}

\subsection*{Result \& Discussion}
\noindent As shown in \cref{tab:video-nofile}, the 5G-NTN link achieves a consistent TTFF of 54--60s, while DVB-S2/RCS2 requires 114--115s, nearly double. The gap stems from uplink access and delivery mechanisms. In DVB-S2/RCS2, the \textit{scheduled return link} forces HTTP GET requests to wait for terminal polling cycles, whereas in OAI’s 5G-NTN, the UE uses RRC and MAC procedures for faster grant-based access. Moreover, 5G-NTN employs native IP tunneling via PDCP and GTP-U \cite{dahlman20205g}, reducing latency, while DVB’s GSE encapsulation adds serialization and buffering delays. TCP connect and start-transfer times are similar ($\approx$0.52s and $\approx$1.06s), but DVB’s higher TTFF reflects slower payload delivery. With both systems at their lowest spectral efficiency, DVB’s framing and low ModCod further prolong startup compared to the more responsive 5G-NTN stack.

\subsection{Experiment-3: Webpage Loading Time}
\noindent This test measures webpage loading performance by recording the time taken to fetch an HTML file from a remote server. Using \texttt{curl} bound to the appropriate interface, a script captures key timing metrics —\texttt{TTFB}, \texttt{time\_connect}, and \texttt{time\_starttransfer}. These metrics help evaluate page responsiveness and network latency from the user's perspective.

\begin{table}[ht]
\centering
\caption{Webpage Load Time}
\label{tab:webpageload}
\begin{tabular}{|>{\centering\arraybackslash}p{0.2cm}
                |>{\centering\arraybackslash}p{0.9cm}
                |>{\centering\arraybackslash}p{0.98cm}
                |>{\centering\arraybackslash}p{0.9cm}
                |>{\centering\arraybackslash}p{0.9cm}
                |>{\centering\arraybackslash}p{0.95cm}
                |>{\centering\arraybackslash}p{0.9cm}|}
\hline
\textbf{\#} & \multicolumn{3}{c|}{\textbf{5G-NTN (s)}} & \multicolumn{3}{c|}{\textbf{DVB-S2/RCS2 (s)}} \\
\hline
& TTFB & Connect & Start Transfer & TTFB & Connect & Start Transfer \\
\hline
1 & 8.10 & 0.53 & 1.09 & 8.09 & 0.54 & 1.07 \\
\hline
2 & 7.92 & 0.52 & 1.06 & 8.12 & 0.54 & 1.07 \\
\hline
3 & 7.92 & 0.52 & 1.06 & 8.12 & 0.54 & 1.06 \\
\hline
4 & 7.90 & 0.52 & 1.05 & 9.23 & 0.53 & 1.06 \\
\hline
5 & 7.96 & 0.52 & 1.05 & 8.02 & 0.53 & 1.05 \\
\hline
\end{tabular}
\end{table}

\subsection*{Results \& Discussion}
\noindent \Cref{tab:webpageload} shows webpage load times for both links. 5G-NTN records TTFB of 7.9--8.1s, while DVB-S2/RCS2 gives 8.0--9.2s; connection and StartTransfer times were nearly identical. Unlike earlier jitter and video TTFF results, the similarity arises because the HTML content was lightweight ($\approx$ 3MB), making load time dominated by TCP setup, HTTP headers, and the fixed 520~ms RTT rather than throughput. Webpage retrieval involves short, single-request flows with minimal packets, and both OAI and OpenSAND employ application-layer buffering to smooth minor queuing delays. These results indicate that for delay-tolerant, small-payload services (e.g., simple webpages, API calls, telemetry polling), 5G-NTN and DVB-S2/RCS2 perform comparably.

\subsection{Experiment-4: File Download}
This test evaluates the effective time experienced by the UE/User-Terminal during file downloads. Using \texttt{curl}, a test file of size 100~MB is downloaded from the application server. \texttt{curl} records file transfer speed. This method offers a practical measure of file transfer performance under real network conditions, relevant to common use cases like updates and content downloads.

\begin{table}[ht]
\centering
\caption{100MB File Download Throughput}
\label{tab:throughput}
\begin{tabular}{|c|c|c|c|}
\hline
\textbf{\#} & \textbf{5G-NTN (kB/s)} & \textbf{DVB-S2/RCS2 (kB/s)} & \textbf{5G / DVB Ratio} \\
\hline
1 & 608 & 274 & 2.22 \\
\hline
2 & 608 & 274 & 2.22 \\
\hline
3 & 616 & 273 & 2.26 \\
\hline
4 & 613 & 274 & 2.24 \\
\hline
5 & 606 & 274 & 2.21 \\
\hline
\end{tabular}
\end{table}

\subsection*{Results \& Discussion}
\noindent As shown in \cref{tab:throughput}, file download throughput over 5G-NTN (OAI) is $\approx$ 610kB/s, more than double that of DVB-S2/RCS2 (OpenSAND) at $\approx$ 274kB/s, despite identical bandwidth (5MHz), RTT (520ms), and low spectral efficiency settings (MCS-1 vs. ModCod-1). With HARQ and RLC-AM disabled in both setups\cite{kumar2023experimental}, 5G-NTN still sustained higher throughput due to slot-based scheduling that enables more frequent transmissions and faster TCP feedback. By contrast, DVB-S2/RCS2’s scheduled return link delays ACKs, slowing congestion window growth. Protocol overhead also differs: 5G-NTN uses leaner encapsulation (GTP-U/PDCP), while DVB’s GSE and framing add latency and reduce efficiency. At low ModCod, DVB suffers from larger frame sizes and wider inter-frame gaps, whereas 5G-NTN maintains shorter, continuous transmissions, yielding superior throughput.

To summarize the practical implications of our findings, \cref{tab:kpi_recommendation} offers a KPI-based recommendation on when each waveform is better suited for specific application types.       
\begin{table}[ht]
\centering
\caption{Recommendation Based on KPI Evaluation}
\label{tab:kpi_recommendation}
\begin{tabular}{|p{4.5cm}|c|c|}
\hline
\textbf{Application Type} & \textbf{5G-NTN} & \textbf{DVB-S2/RCS2} \\
\hline
Jitter-sensitive applications (e.g., VoIP) & \checkmark & \\
\hline
Interactive file transfers (e.g., HTTP GET, updates) & \checkmark & \\
\hline
Webpage loading (light HTML) & \checkmark & \checkmark \\
\hline
Video streaming & \checkmark & \\
\hline
Sustained download throughput & \checkmark & \\
\hline
\end{tabular}
\end{table}

\section{Conclusion}\label{sec:conclusion}
\noindent This paper presented a user-centric application-layer comparison of 5G-NTN and DVB-S2/RCS2 using open-source platforms: OAI 5G-NTN and OpenSAND, within the 6G-SANDBOX testbed. Unlike prior works focused on lower-layer metrics, our study evaluated real-world KPIs such as jitter, file download time, and video/webpage load latency. Results show that 5G-NTN consistently offers lower jitter and faster access, thanks to its flexible scheduling and leaner protocol stack. DVB-S2/RCS2, while mature and efficient for scheduled delivery, exhibits higher delays due to its TDM framing and return link structure. Our findings highlight the importance of open, reproducible tools for NTN research and provide early insights into protocol-level trade-offs. Future work will address more complex traffic patterns, mobility, and multi-user scenarios.

\section*{Acknowledgment}
This work was carried out as part of the project \textit{NTN-WAVE}, supported by the European Union’s Horizon Europe program under the 6G-SANDBOX project (Grant Agreement No.~101096328).

\printbibliography

@book{dahlman20205g,
  title={5G NR: The next generation wireless access technology},
  author={Dahlman, Erik and Parkvall, Stefan and Skold, Johan},
  year={2020},
  publisher={Academic Press}
}

@techreport{etsiTR103886,
  author       = {{ETSI TC SES}},
  title        = {{Satellite Earth Stations \& Systems (SES); DVB-S2x/RCS2 versus 3GPP New Radio protocol technical comparison for broadband satellite systems}},
  institution  = {ETSI},
  type         = {Technical Report},
  number       = {TR 103 886},
  note         = {Unpublished report, V0.0.9 Draft},
  month        = {10},
  year         = {2024}
}

@article{huikko2023comparison,
  title={Comparison of 5G NR and DVB-S2X/RCS2 Technologies in Broad-Band Satellite Systems},
  author={Huikko, Tuomas},
  journal={PhD, University of Tampere},
  year={2023}
}

@inproceedings{delbeke2022satellite,
  author    = {Philippe Delbeke},
  title     = {Satellite Return Waveform Efficiency Comparison for 5G-NR, DVB-S2X, DVB-RCS2 and Mx-DMA MRC},
  booktitle = {Proceedings of the 27th Ka Broadband Communications Conference},
  address   = {Stresa, Italy},
  date      = {2022-10-18},
  year      = {2022},
  pages     = {},
  note      = {October 18--21, 2022}
}

@misc{openairinterface5g_rfsimulator,
  author       = {{OpenAirInterface Software Alliance}},
  title        = {OpenAirInterface5G RF Simulator Source Code: \texttt{simulator.c}},
  howpublished = {\url{https://gitlab.eurecom.fr/oai/openairinterface5g/-/blob/develop/radio/rfsimulator/simulator.c}},
  year         = {2024},
  note         = {Accessed: 2025-05-25}
}

@inproceedings{sormunen2025simulative,
  title={Simulative Comparison of DVB-S2X/RCS2 and 3GPP 5G NR NTN Technologies in a Geostationary Satellite Scenario},
  author={Sormunen, Lauri and Huikko, Tuomas and R{\"o}nty, Verneri and Sepp{\"a}nen, Erno and Rantanen, Sami and Laakso, Frans and Hyt{\"o}nen, Vesa and Majamaa, Mikko and Puttonen, Jani},
  booktitle={2025 12th Advanced Satellite Multimedia Systems Conference and the 18th Signal Processing for Space Communications Workshop (ASMS/SPSC)},
  pages={1--8},
  year={2025},
  organization={IEEE}
}

@inproceedings{george20215g,
  title={5G {N}ew {R}adio in {N}onlinear {S}atellite {D}ownlink: A physical layer comparison with DVB-S2X},
  author={George, Geordie and Roy, Samhita and Raghunandan, Sahana and Rohde, Christian and Heyn, Thomas},
  booktitle={2021 IEEE 4th 5G World Forum (5GWF)},
  pages={499--504},
  year={2021},
  organization={IEEE}
}

@inproceedings{kumar2023experimental,
  title={Experimental study of the effects of RLC modes for 5G-NTN applications using OpenAirInterface5G},
  author={Kumar, Sumit and Sheemar, Chandan Kumar and Querol, Jorge and Nik, Amirhossein and Chatzinotas, Symeon},
  booktitle={2023 IEEE Globecom Workshops (GC Wkshps)},
  pages={233--238},
  year={2023},
  organization={IEEE}
}

@misc{opensand,
  author       = {{OpenSAND}},
  title        = {OpenSAND - Open Satellite Network Emulator},
  howpublished = {\url{https://www.opensand.org/}},
  note         = {Accessed: 2025-05-25}
}

@inproceedings{kumar2022openairinterface,
  title={OpenAirInterface as a platform for 5G-NTN Research and Experimentation},
  author={Kumar, Sumit and Meshram, Ashish Kumar and Astro, Abdelrahman and Querol, Jorge and Schlichter, Thomas and Casati, Guido and Heyn, Thomas and V{\"o}lk, Florian and Schwarz, Robert T and Knopp, Andreas and others},
  booktitle={2022 IEEE Future Networks World Forum (FNWF)},
  pages={500--506},
  year={2022},
  organization={IEEE}
}

@misc{6gsandbox,
  author       = {{6G Sandbox}},
  title        = {6G Sandbox - Open 6G Research Infrastructure},
  howpublished = {\url{https://6g-sandbox.eu/}},
  note         = {Accessed: 2025-05-25}
}

@misc{opensand_link_budget,
  author       = {{openSAND}},
  title        = {OpenSAND Link Budget},
  howpublished = {\url{https://github.com/CNES/opensand/wiki/link-budget}},
  note         = {Accessed: 2025-05-25}
}

@inproceedings{kumar20225g,
  title={5G-NTN GEO-based over-the-air demonstrator using OpenAirInterface},
  author={Kumar, Sumit and Kodheli, Oltjon and Astro, Abdelrahman and Querol, Jorge and Chatzinotas, Symeon and Casati, Guido and Schlichter, Thomas and Heyn, Thomas and Cheporniuk, Hlib and V{\"o}lk, Florian and others},
  booktitle={39th International Communications Satellite Systems Conference (ICSSC 2022)},
  volume={2022},
  pages={110--114},
  year={2022},
  organization={IET}
}

@article{volk20245g,
  title={5G non-terrestrial networks with OpenAirInterface: An experimental study over GEO satellites},
  author={V{\"o}lk, Florian and Schlichter, Thomas and Kumar, Sumit and Schwarz, Robert T and Knopp, Andreas and Hammouda, Marwan and Heyn, Thomas and Querol, Jorge and Chatzinotas, Symeon and Kapovits, Adam},
  journal={IEEE Access},
  year={2024},
  publisher={IEEE}
}

@inproceedings{kumar20235g,
  title={5G NTN LEO based demonstrator using OpenAirInterface5G},
  author={Kumar, Sumit and Sheemar, Chandan Kumar and Querol, Jorge and Yilmaz, Turker and Chatzinotas, Symeon and Hammouda, Marwan and Heyn, Thomas and Schlichter, Thomas and Marques, Paulo and Pereira, Luis and others},
  booktitle={IET Conference Proceedings CP873},
  volume={2023},
  number={48},
  pages={69--75},
  year={2023},
  organization={IET}
}

@inproceedings{kumar20225ag,
  title={5G-NTN GEO-based in-lab demonstrator using OpenAirInterface5G},
  author={Kumar, Sumit and Abdalla, Abdelrahman and Kodheli, Oltjon and Querol, Jorge and Chatzinotas, Symeon and Schlichter, Thomas and Casati, Guido and Heyn, Thomas and Volk, Florian and Kaya, Sertac and others},
  booktitle={11th Advanced Satellite Multimedia Conference},
  year={2022}
}

\end{document}